# A Modified Kwee–Van Woerden Method for Eclipse Minimum Timing with Reliable Error Estimates

Hans J. Deeg [1,2]

1 Instituto de Astrofísica de Canarias, 38205 La Laguna, Tenerife, Spain; hdeeg@iac.es
2 Department de Astrofísica, Universidad de La Laguna, 38206 La Laguna, Tenerife, Spain

**Abstract:** The Kwee–van Woerden (KvW) method used for the determination of eclipse minimum times has been a staple in eclipsing binary research for decades, due its simplicity and the independence of external input parameters, which also makes it well-suited to obtaining timings of exoplanet transits. However, its estimates of the timing error have been known to have a low reliability. During the analysis of very precise photometry of CM Draconis eclipses from TESS space mission data, KvW's original equation for the timing error estimate produced numerical errors, which evidenced a fundamental problem in this equation. This contribution introduces an improved approach for calculating the timing error with the KvW method. A code that implements this improved method, together with several further updates of the original method, are presented. An example of the application to CM Draconis light curves from TESS is given. The eclipse minimum times are derived with the KvW method's three original light curve folds, but also with five and seven folds. The use of five or more folds produces minimum timings with a substantially better precision. The improved method of error calculation delivers consistent timing errors which are in excellent agreement with error estimates obtained by other means. In the case of TESS data from CM Draconis, minimum times with an average precision of 1.1 seconds are obtained. Reliable timing errors are also a valuable indicator for evaluating if a given scatter in an O-C diagram is caused by measurement errors or by a physical period variation.

**Keywords:** eclipsing binary minima timing method; transit timing variation method; eclipsing binary stars; exoplanet transits; individual stars (CM Draconis); TESS space mission; computational methods

## 1. Introduction

The Kwee–van Woerden method (KvW) has been very popular for eclipse minimum time determination since its publication in 1956 [1]. This is due to its computational simplicity and due to its independence from assumptions about the data that are being analyzed, beyond the assumption of data points being equally spaced over time, with a symmetric eclipse shape. However, practitioners have long been aware of the unreliability of the algorithm's error estimates, which are typically considered as being too optimistic [2–4]. During the analysis of highly precise eclipse time-series from the TESS space mission, KvW's equation for the timing error estimate went however from unreliable to unsolvable, which motivated the modification of the error estimate described in this paper. We note that reliable timing errors are a valuable indicator for evaluating if a given scatter in an O-C diagram is caused by measurement errors or by a physical period variation.

The KvW method, in brief, assumes a light curve of an eclipse of *N* equidistant points separated by $\Delta t$, in which a given data point at time $T_1$ represents a preliminary minimum time. Using $T_1$ as the reflection axis, the differences in the magnitudes or fluxes between the paired points $\Delta m_k$ ($k = 1...n$) are taken, and their squared sum is calculated as $S(T_1) \equiv \sum$ $(\Delta m_k)^2$. The symmetry axis is then shifted to $(T_1 - \frac{1}{2}\Delta t)$ and $(T_1 + \frac{1}{2}\Delta t)$, and the corresponding sums of $S(T_1 - \frac{1}{2}\Delta t)$ and $S(T_1 + \frac{1}{2}\Delta t)$ are calculated while keeping the number of pairings, $n$, the same in all reflections. The values of $S$ against time are then fit by a parabola of the form

$$S_{fit}(T) = a\,T^2 + bT + c. \qquad (1)$$

This parabola has a minimum value of $S_{fit}(T_0)$ given by

$$S_{fit}(T_0) = c - b^2/4a \qquad (2)$$

at the time

$$T_0 = -b/2a, \qquad (3)$$

which is the sought-after minimum time. For ascertaining the error of the minimum time, $\sigma_{T0}$, KvW employs the following equation:

$$\sigma^2_{T0} = (4ac - b^2)/(4a^2\,(Z-1)), \qquad (4)$$

where $Z$ is the maximum number of independent flux pairings, with $Z = N/2$ in the case of equidistant points. It should be noted that the original KvW uses only three reflections for the calculation of $S$. Later implementations may also use five or seven reflections spaced by further $\frac{1}{2}\Delta t$-steps away from $T_1$, which we call 3-, 5-, and 7-fold implementations of the KvW.

## 2. Identification of the Problem

The failure of Equation (4) became apparent when we employed the original 3- or 5-fold KvW to determine $T_0$ on individual eclipses of the well-characterized M4-M4 binary CM Dra [5] in very precise light curves from the TESS space mission [6]. Individual TESS light curves have lengths of about 28 days, and therefore contain 17–19 primary as well as secondary eclipses of CM Dra, which has a period of 1.268 days, with primary and secondary eclipses of rather similar depths of 47.5% and 44.5%, respectively.

In several individual eclipses, a computational error arose when attempting to solve Equation (4) for $\sigma_{T0}$, caused by taking a root with a negative value of the term $4ac - b^2$. This condition of $4ac - b^2 < 0$ is equivalent to the minimum value of $S_{fit}$ (Equation (2)) becoming negative (see also Figure 1).

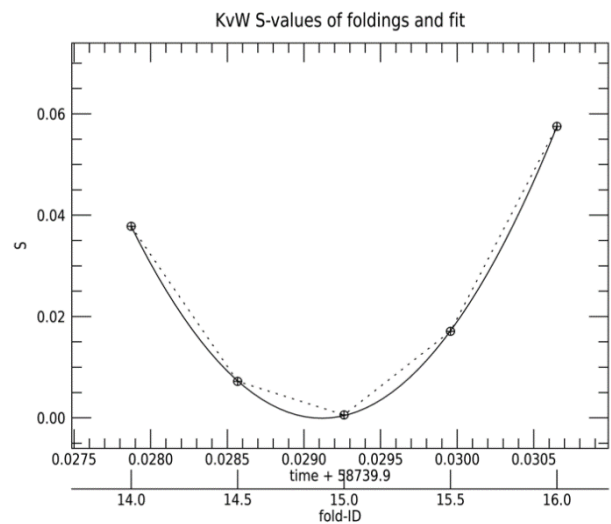

**Figure 1.** *S* values (sum of squared differences between fluxes in pairings) for the first complete primary eclipse of CM Dra observed by TESS (see also Fig. 2). The upper X axis shows the time in units of BJD-2400000 and the lower one gives the enumeration ('fold-ID') of the flux-values on which (or between which) the folding was performed. The circles are *S* values from five folds at the given fold-ID, while the solid line is the second-order polynomial that is fitted through these points. It should be noted that the minimum of the fitted curve is slightly below zero.

While the derivation of $\sigma_{T0}$ from Equation (4) evidently failed in some cases, it should also be noted that $\sigma_{T0}$ may approach zero—and potentially be underestimated by orders of magnitude—whenever a positive term of $4ac - b^2$ approaches zero.

Tests were then performed with TESS light curves with artificially added noise. In these cases, the minimum values $S_{fit}(T_0)$ increased and the numerical problems in the determination of $\sigma_{T0}$ vanished. Therefore, problems in the determination of $\sigma_{T0}$, as well as serious underestimations of $\sigma_{T0}$ from very small values of $4ac - b$, have been a consequence of the significant increase in instrumental photometric precision in the 65 years since the algorithm's publication.

### 3. A Revised Determination of the Timing Error

Considering the average noise of the flux-measurements to be $\mu = \langle\mu_i\rangle$, with $\mu_i$ being the observational error of an individual measurement, the average noise in the difference of two fluxes, $\Delta m_k = m_{+k} - m_{-k}$, is given by $\mu\sqrt{2}$.

Instead of using $S(T_j)$, with $T_j$ being the epoch of the reflection axis, we may also calculate the usual $\mathcal{X}^2$ statistical values for these pairings, which are

$$\mathcal{X}^2(T_j) = RSS/\mu^2 = S(T_j)/2\mu^2, \qquad (5)$$

where $RSS$ is the residual sum square, given by $S(T_j)/2$. If we assume that the flux from the observed binary is perfectly symmetric around the minimum time $T_0$, then a corresponding minimum value of $S_{fit}(T)$ at $T = T_0$ should be fully dominated by observational errors. This corresponds to KvW's Equation (13) [3] of $S(T_0) = \sum (\mu_{+k} + \mu_{-k})^2$, where $\mu_{\pm k}$ represents the deviations in magnitude at the time $T_0 \pm k\,\Delta t$. In the same situation, when the measurement error dominates, the minimum value of $\mathcal{X}^2(T)$ at $T = T_0$ is determined by the number of degrees of freedom $Z$, given by

$$\mathcal{X}^2(T_0) = Z - 1. \qquad (6)$$

Following Equation (5), the equivalent expression for the value $S(T_0)$ is then

$$S(T_0) = (Z - 1)\,2\mu^2. \qquad (7)$$

It should be noted that KvW's Equation (13) [3] also indicates that $S(T_0)$ should have this value, except for a factor $Z/(Z - 1)$ that is close to unity. In practice, however, $S(T_0)$ may have values that substantially deviate from Equation (7) or from KvW's Equation (13) [3], which is the origin of the poor error estimates of the original KvW method.

In our modified KvW algorithm, after an initial determination of the coefficients $a$, $b$, and $c$ as usual, we therefore offset the fitted $S_{fit}(T_0)$ so that $S_{fit}(T_0)$ is *defined* by Equation (7). Using Equation (2), the coefficient $c$ is then determined so that $S_{fit}(T_0)$ complies with Equation (7), with $c$ being given by

$$c = (Z-1)\,2\mu^2 + b^2/4a. \qquad (8)$$

Inserting the revised value of $c$ from Equation (8) into KvW's original error determination, Equation (4) is simplified to

$$\sigma_{T0}^2 = 2\mu^2/a. \qquad (9)$$

The average flux error $\mu$ should preferentially be supplied as an external parameter, from a measurement of the time-series' noise outside of the eclipses. If this is not possible, as an alternative, we may assume that the lowest $S(T)$ obtained from folds on or between data points is dominated by the noise $\mu$, while contributions to that $S(T)$ from remaining imperfections in the fold's symmetry are relatively small. $\mu$ can then be derived from an inversion of Equation (7). In practical cases, if at least two data points are in an eclipse's central flat part, this method led to values of $\mu$ that are within 50% of a $\mu$ measured from the off-eclipse noise.

It should be noted that a conversion to $\mathcal{X}^2$ statistics by fitting a parabola to the $\mathcal{X}^2(T_j)$ instead of the $S(T_j)$, while using the interrelation from Equation (5), leads to $\sigma_{T0} = 1/\sqrt{a}$. This corresponds to the usual definition of the 1-$\sigma$ region of confidence for one free parameter, where a quadratic function at the points $\mathcal{X}^2(T_0 \pm \sigma_{T0})$ is increased by 1 over the minimum value $\mathcal{X}^2(T_0)$.

## 4. Code Implementation of the Kwee–van Woerden Method with Improved Error Estimates

A code named kvw.pro has been written in the IDL language that implements the KvW method with the timing error estimate as described above (see the Supplementary Materials for a link to the code). It is not overly complex and is heavily commented on. A core version (kvwcore.pro) that is stripped of non-essential output options has also been made available, in order to facilitate its implementation in other languages. The code provides several further improvements over a basic implementation of the KvW method, which are itemized in the following:

- Test for equidistance of input flux points: Similar to the original KvW method, the code requires data points that are equally spaced over time. The code tests whether variations in temporal spacing larger then 1% of the median spacing occur, and if so, whether they halt further processing. If this is considered too stringent, the rejection value can be modified. If needed, data input to kvw.pro should be converted into equidistant flux-points through prior linear (as proposed in the original paper by KvW) or higher-order interpolations or fits;
- Selection of data points: While the user has to take care that an input time-series only contains data collected during an eclipse (usually requiring a minimum flux-drop against the off-eclipse flux, see Figure 2), asymmetric data coverage around the center of an eclipse is recognized by the code. The code always selects the maximum number of data points that are available for pairings on either branch of an eclipse, and hence balances the coverage between the ingress and the egress (Figure 3);
- Employment of more than three folds around the initial minimum time estimate: The number of folds needs to be odd and the use of five (default) or seven folds is recommended;
- For the initial minimum time estimate, the algorithm uses—by default—the central point of the supplied light curve, but the user may also choose to use the point with the lowest flux. The central value is the better choice unless there are considerable asymmetries in the eclipse light curve. The point of the least flux should only be used in low-noise data, when this point is well-defined against the noise of the curve;
- The code automatically selects the maximum amount of data points that can be paired for folds, which avoids errors when data of incomplete eclipses are provided (see Figure 3).
- Symmetrizing the fit to $S(T)$: If the lowest value of $S(T)$ does not correspond to a fold that is close to the initial estimate of the minimum time, the fitted curve $S_{fit}$ will have branches of unequal length and the longer branch either to the left or right of the lowest $S(T)$ has a larger weight for the coefficients *a*, *b*, and *c*. The same situation may also arise when data from incomplete eclipses are analyzed (see also Figure 4). As a remedy, the outer values of $S(T)$ for the longer branch are cropped, so that this branch is at most one point larger than the shorter one. The fit for $S_{fit}$ is then performed based on the reduced set of values $S(T)$. This cropping can only be performed if $S(T)$ has been obtained at more than three folds;
- The code also permits a determination of the timing error using the original procedure of KvW from 1956, which does not require an explicit determination of the noise of the flux values;

- Optionally, a graphical output similar to Figure 1, 3, and 4 can be produced, which may provide useful diagnostics for the revision of the timing measurements.

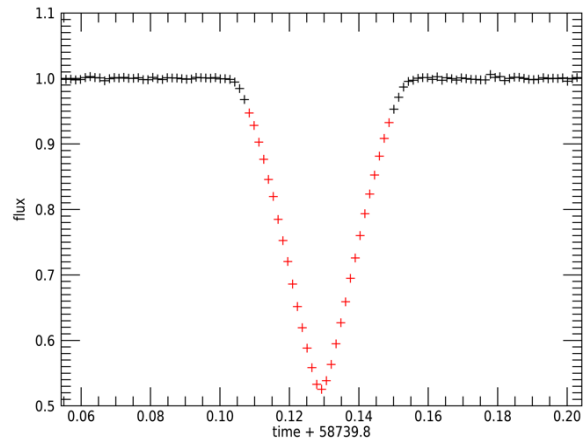

**Figure 2.** Light curve of the first complete primary eclipse in the TESS data, at BJD 2458739.9291. The y-axis represents flux values that are normalized to the off-eclipse flux using the procedure described in the text and the x axis indicates BJD-2400000. Points shown in red have a flux of <0.95, which are those that were used as input for the Kwee–van Woerden (KvW) algorithm.

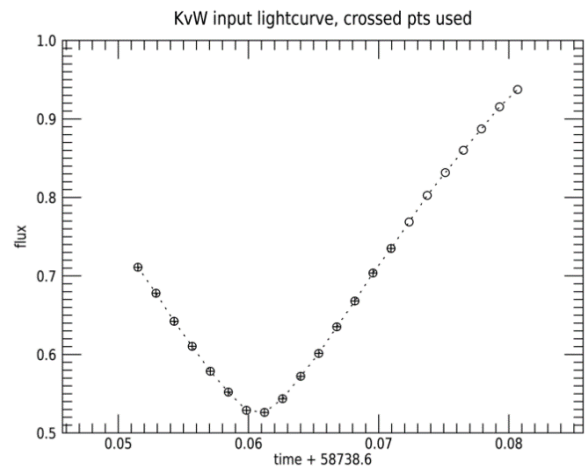

**Figure 3.** Input light curve for the KvW algorithm of the first—but incomplete—CM Dra eclipse in TESS data, which is a primary eclipse at BJD 2458738.6607. The y axis represents the normalized flux and the x axis indicates BJD-2400000. The code selects the maximum amount of data points that can be paired for folds (filled circles) and ignores the others (open circles). When numbering the points by counting from zero, the eclipse minimum is between points 6 and 7 (see the 'fold-ID' in Figure 4).

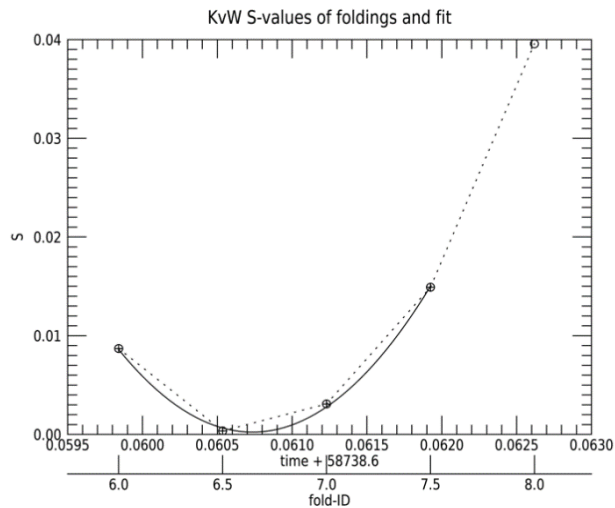

**Figure 4.** Similar to Figure 1, but showing the *S*-values from the incomplete eclipse of Figure 3. In this case, the distribution of *S*-values is asymmetric around the lowest *S*-value at fold-ID 6.5. The code then rejects the right-most fold (open circle) and performs a fit based on only the four remaining S values.

## 5. Example Application to TESS Data and Verification of the Error Estimates

In the following section, we present the application of the modified KvW to the aforementioned TESS data of CM Dra. The analyzed data were the first which TESS obtained on CM Dra, in TESS Sector 16, which were acquired from 11 November 2019 to 7 October 2019. In addition to Sector 16, CM Dra was also observed in TESS sectors 19 and 22 to 26; however, their analysis is beyond the scope of this publication. The light curve was downloaded from NASA's MAST archive, and had previously been processed with its pipeline version spoc-4.0.28 (We note that light curves available on MAST until Spring 2020, which had been processed by versions prior to spoc-4.0.26, had time-stamps that were 2 s too large [7]). From this dataset, we used the 'PDCSAP_FLUX' values, which are fluxes that underwent a 'Pre-search Data Conditioning' procedure to remove common instrumental effects [8]. Around each individual eclipse, a time-series covering about three times the eclipse duration of 0.050 d was extracted; see Figure 2 for an example of an extracted section. For each eclipse snippet, a second-order polynomial was then fitted to the off-eclipse sections before and after the eclipse. The fluxes were then divided by the polynomial fit, resulting in an eclipse light curve whose off-eclipse flux was normalized to 1 and which was free of gradients and other signals on time-scales larger than a few hours, be they from CM Dra or from instrumental effects (again see Figure 2).

The flux error $\mu$ was determined from the off-eclipse point-to-point *rms* of these normalized light curves. Among individual eclipses, it varied between $0.98 \times 10^{-3}$ and $2.45 \times 10^{-3}$ in normalized flux units. Since the higher of these *rms* values were dominated by individual flux-peaks in the off-eclipse data, for the further analysis, we used a noise value that was averaged over all eclipses, which was $\mu = 1.38 \times 10^{-3}$ or 1380 ppm. The modified KvW with the default of five folds was then executed separately on each of the primary and secondary eclipses, using only data points with relative fluxes of less than 0.95 (red points in Figure 2).

The resulting minimum times and errors of all complete eclipses—18 primary and 18 secondary ones—are shown in Table 1, while Figure 5 presents their observed minus calculated (O-C) values against the ephemeris given by [9] (in [9], the epochs of reference are given in BJD_TAI. These have been converted to BJD_TDB by adding 32.184 s). For comparison, the timing error from the original KvW method is also given in Table 1. Its calculation resulted in numerical errors (indicated as NaN) in half of the eclipses, due to the negative root mentioned previously. The NaN values occurred whenever the algorithm calculated the polynomial fit on all five points of *S*, whereas the eclipses in which the algorithm decided to symmetrize the fit to *S* by fitting over four points only—similar to the situation of Figure 4—resulted in finite values of $\sigma_{T0}$. It should be noted that an original implementation of the KvW method would not contain the symmetrization, so we may expect that it would result in NaNs for even more of the eclipses.

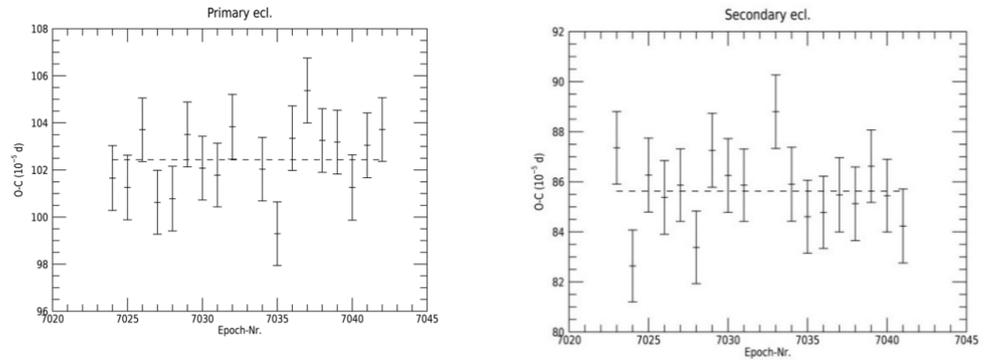

**Figure 5.** Observed minus calculated (O-C) eclipse minimum times for CM Dra primary and secondary eclipses in TESS sector 26 data. The observed times correspond to Table 1 and the calculated times and epoch numbers are from the ephemeris of [9]. The dashed line is the average O-C value of these eclipses. The size of the error bars is in near perfect agreement with a Gaussian distribution of the minimum times around the dashed line, where the distribution's width corresponds to the size of the error bars.

**Table 1.** CM Dra eclipse minimum times from the KvW method with five folds, with timing errors from the revised error estimate ($\sigma_{T0}$) and from KvW's original formula ($\sigma_{T0KvW}$), from TESS Sector 26 observations. Epoch numbers are relative to BJD 49830.757712 for primary eclipses and BJD 2549831.390742 for secondary eclipses.

| Epoch Nr. | $T_0$ | $\sigma_{T0}$ | $\sigma_{T0KvW}$ |
|---|---|---|---|
| | (BJD-TBD-2400000) | ($10^{-5}$ d) | ($10^{-5}$ d) |
| | Primary eclipses | | |
| 7024 | 58739.9291169 | 1.25 | NaN |
| 7025 | 58741.1975015 | 1.25 | NaN |
| 7026 | 58742.4659011 | 1.26 | 2.78 |
| 7027 | 58743.7342864 | 1.26 | 2.20 |
| 7028 | 58745.0026702 | 1.25 | NaN |
| 7029 | 58746.2710774 | 1.25 | NaN |
| 7030 | 58747.5394452 | 1.26 | 2.81 |
| 7031 | 58748.8078623 | 1.26 | 2.19 |
| 7032 | 58750.0762635 | 1.25 | NaN |
| 7034 | 58752.6130101 | 1.26 | 2.85 |
| 7035 | 58753.8813997 | 1.25 | 2.06 |
| 7036 | 58755.1498107 | 1.25 | NaN |
| 7037 | 58756.4182122 | 1.25 | NaN |
| 7038 | 58757.6865806 | 1.27 | 2.81 |
| 7039 | 58758.9549907 | 1.26 | 1.90 |
| 7040 | 58760.2233508 | 1.26 | NaN |
| 7041 | 58761.4917508 | 1.24 | NaN |
| 7042 | 58762.7601436 | 1.26 | 2.86 |
| Mean | | 1.25 ± 0.01 | 2.50 ± 0.40 |

| | Secondary eclipses | | |
|---|---|---|---|
| 7023 | 58739.2935944 | 1.34 | 2.23 |
| 7024 | 58740.5619511 | 1.36 | 3.23 |
| 7025 | 58741.8303857 | 1.34 | NaN |
| 7026 | 58743.0987637 | 1.33 | NaN |
| 7027 | 58744.3671428 | 1.35 | 2.18 |
| 7028 | 58745.6355326 | 1.35 | 2.95 |
| 7029 | 58746.9039510 | 1.33 | NaN |
| 7030 | 58748.1723260 | 1.33 | NaN |
| 7031 | 58749.4407154 | 1.36 | 3.78 |
| 7033 | 58751.9775280 | 1.33 | NaN |
| 7034 | 58753.2458785 | 1.33 | NaN |
| 7035 | 58754.5142489 | 1.36 | 2.34 |
| 7036 | 58755.7826651 | 1.34 | 2.75 |
| 7037 | 58757.0510572 | 1.34 | NaN |
| 7038 | 58758.3194374 | 1.34 | NaN |
| 7039 | 58759.5878263 | 1.35 | 2.32 |
| 7040 | 58760.8562353 | 1.35 | 2.58 |
| 7041 | 58762.1246012 | 1.33 | NaN |
| Mean | | 1.34 ± 0.01 | 2.71 ± 0.53 |

Based on Table 1, the average of the individual timing errors from the revised calculation is $1.25 \times 10^{-5}$ d for primary minima, and $1.34 \times 10^{-5}$ d for secondary minima (or 1.08 and 1.16 s, respectively), with individual errors only varying within $\sim 10^{-7}$ d around these averages. These timing errors and further timing errors discussed in this paper are compiled in Table 2.

An independent estimate of the timing error could be obtained from the standard deviation of the O-C values against the mean O-C value (dashed lines in Figure 5). This value is $1.18 \times 10^{-5}$ d for primary eclipses and $1.48 \times 10^{-5}$ d for secondary eclipses. This is in very good agreement with the above-mentioned errors, and shows that the timing errors from individual eclipse timings are reliable.

Further independent verification could be obtained from the timing error estimator (TEE) equations presented by [10]. The referenced work provides several equivalent formulae for the timing error, based on a light curve's noise and data cadence, and on the eclipse depth and duration. Here, we use their Equation (7), given as

$$\sigma_{T0} = \mu\, T_\nabla / (2\, \Delta F\, \sqrt{n_\nabla}), \qquad (10)$$

where $\Delta F$ is the relative depth of the eclipse, $T_\nabla$ is the combined in- and egress duration, and $n_\nabla$ is the number of data points within $T_\nabla$. For the photometric noise, we used again $\mu = 1.38 \times 10^{-3}$ and for $\Delta F$, we employed the above-mentioned eclipse depths of 0.475 and 0.445. The duration was determined as $T_\nabla = 0.050$d, which corresponds to $n_\nabla = 36$ data points, given the 2-min cadence of TESS. With Equation (10), we obtained timing errors of $1.21 \times 10^{-5}$ d for primary eclipses and $1.29 \times 10^{-5}$ d for secondary eclipses, which is again in excellent agreement with the results obtained by the modified KvW.

Repetition of the above analysis using only three folds, as in KvW's original implementation, led to a notably larger scatter of the measured minimum times against the mean O-C value, with values of $1.60 \times 10^{-5}$ d and $1.65 \times 10^{-5}$ d for primary and secondary eclipses, respectively, while the revised method's timing errors remained essentially unchanged against the use of five folds (see Table 2). The original KvW error determination resulted in two eclipses with numerical errors, but produced timing errors whose average agrees well with the scatter around the mean O-C value, albeit with a large variation among error estimates from individual eclipses, ranging from $0.85 \times 10^{-5}$ d to $2.83 \times 10^{-5}$ d (see Appendix A, Table A1). There were no perceivable differences in the quality of the light curves among the different eclipses. The use of seven folds, on the other hand, led to corresponding scatters against the mean O-C value of $1.28 \times 10^{-5}$ d and $1.36 \times 10^{-5}$ d, which is in near-perfect agreement with the average size of the individual eclipses' timing errors, whereas the original KvW error estimates led to substantially larger values, albeit without any numerical errors (Appendix A, Table A2).

**Table 2.** Timing precision of CM Dra eclipses obtained by different methods discussed in the text, for primary and secondary eclipses. Values explicitly mentioned in the text are presented in bold.

| Description | $\sigma_{T0,prim}$ | $\sigma_{T0,sec}$ |
|---|---|---|
| | ($10^{-5}$ d) | ($10^{-5}$ d) |
| Three-fold KvW method | | |
| Timing errors of individual eclipses, KvW original calculation | $1.62 \pm 0.48$ | $1.63 \pm 0.33$ [1] |
| Ditto, revised calculation | $1.24 \pm 0.01$ | $1.32 \pm 0.01$ |
| Standard dev. of minimum times against mean O-C value | **1.60** | **1.65** |
| | | |
| Five-fold KvW method | | |
| Timing errors of individual eclipses, KvW original calculation | $2.50 \pm 0.40$ [2] | $2.71 \pm 0.53$ [2] |
| Ditto, revised calculation | **1.25 ± 0.01** | **1.34 ± 0.01** |
| Standard dev. of minimum times against mean O-C value | **1.18** | **1.48** |
| | | |
| Seven-fold KvW method | | |
| Timing errors of individual eclipses, KvW original calculation | $3.89 \pm 1.79$ | $3.78 \pm 1.71$ |
| Ditto, revised calculation | $1.27 \pm 0.01$ | $1.36 \pm 0.01$ |
| Standard dev. of minimum times against mean O-C value | **1.28** | **1.36** |
| | | |
| Timing Error Estimator (TEE), from [10] | **1.21** | **1.29** |

Notes: NaN values were ignored in the calculation of averages and standard deviations. [1] Two of the 18 eclipses have NaN values. [2] Nine of the 18 eclipses have NaN values.

## 6. Conclusions

A method for calculating reliable minimum time errors using the Kwee–van Worden algorithm has been presented. This updated method only affects the error estimate, while the determination of the minimum time itself proceeds along KvW's original prescription, which continues to be of interest due to the independence of the timing measurement from any further assumption or knowledge of the binary under investigation. However, the associated computer code gives the option to use more than the three time folds of KvW's original presentation. Both KvW's original method and the current code are the most suitable for V- or U-shaped eclipses. In flat-bottomed eclipses, pairings of data points from the flat part only add noise without information to the values of the *S(T)*, and hence degrade the precision of the timing measurements. In principle, the KvW method could be modified to exclude the flat central parts of an eclipse from the pairings which determine *S(T)*, using only the data from the in- and egress. For the sake of simplicity, this has not been implemented in the current version of the code, but remains pending for future updates.

The application of the updated method to TESS light curves of CM Draconis with multiple eclipses demonstrated an excellent agreement between the size of the timing measurement errors and the scatter of the measured minimum times. This agreement was within 25% when using KvW's original three folds, but improved to 5–10% for five folds and was within ~1% for seven folds. With the updated method, the timing errors only varied by ~1% between individual eclipses, and only displayed a small dependency on the number of folds used. In comparison, error estimates from KvW's original equation with three folds only led in average to a good value, but the errors scatter widely among individual eclipses (and two numerical errors occured). The frequency of numerical failures with the original KvW method increased when using five folds, while the remaining error estimates became significantly poorer. For seven folds, no numerical failures occurred in KvW's original error estimates, but the error estimates became even worse. In all levels of folds, error estimates from KvW's original formula exhibited a wide scatter among individual eclipses. The likely explanation for this is their dependence on all three parameters of *a*, *b*, and *c* of the parabolic fit to *S*. The revised error estimate, on the other hand, only depends (see Equation (9)) on the photometric noise $\mu$—which was considered to be identical for all eclipses—and on the quadratic coefficient *a* of the parabolic fit, where *a* turned out to vary little across individual eclipses and with the number of folds.

Considering the scatter of the measured minimum times against the mean O-C value—which gives us an independent measure of the quality of the minimum timing measurements—we observed a clear improvement when using five or seven folds over three folds. The recommended application of the KvW method is therefore the use of five, and preferentially seven, folds together with the revised error estimate presented here. In the case of doubt about the reliability of an error estimate, a comparison with one of the Timing Error Estimator (TEE) equations of [10] is also recommended. An Excel sheet implementing the TEE is accessible through the Supplementary Materials.

For seven folds, we found a near-perfect agreement between the individual eclipses' timing errors and the scatter of the minimum times against the mean O-C value. This also showed that CM Dra's timing deviations against the mean O-C (over the 28-day section that was analyzed) are very well-described by a Gaussian distribution, whose width is given by the individual timing errors. In turn, this implies that the photometric noise of the analyzed TESS data—after the treatment described in Section 5—only has negligible components of non-white noise, at least on the time-scales between the 2-min data cadence (which dominates the individual eclipses' timing error) and the hour-long duration of the binary eclipses (which dominates the scatter in O-C times). In the same vein, the consistency between the scatter around the mean O-C value and the size of the timing error indicates an absence of any physical period variation within the short time span that has been analyzed. The timing precision of about 1.1 s that was obtained for each CM Dra eclipse shows that TESS data contain a rich trove of eclipsing binary data that may be analyzed against previously obtained eclipse timings. Besides the eclipses from TESS Section 16 which have been shown here, TESS has acquired data of CM Dra in Sections 19 and 22–26, whose analysis is the subject of forthcoming work.

**Supplementary Materials:** The code described in this paper is available as open source software at https://github.com/hdeeg/KvW.

**Funding:** This research was funded by the Spanish Research Agency of the Ministry of Science and Innovation (AEI-MICINN) under grants entitled 'Contribution of the IAC to the PLATO Space Mission', with references ESP2017-87676-C5-4-R and PID2019-107061GB-C66, DOI: 10.13039/501100011033.

**Data Availability Statement:** Please refer to suggested Data Availability Statements in section "MDPI Research Data Policies" at https://www.mdpi.com/ethics.

**Acknowledgments:** This paper includes data collected by the TESS mission. Funding for the TESS mission is provided by the NASA Explorer Program. The author thanks the three anonymous referees for their feedback, which led to a significantly improved presentation of this work.

**Conflicts of Interest:** The author declares no conflicts of interest.

## Appendix A. Tables of CM Dra Minimum Times and Errors with Three and Seven Folds

**Table A1.** Similar to Table 1, but using the KvW method with three folds.

| Epoch-Nr. | $T_0$ | $\sigma_{T_0}$ | $\sigma_{T_0 KvW}$ |
| --- | --- | --- | --- |

| | (BJD-TBD-2400000) | (10$^{-5}$ d) | (10$^{-5}$ d) |
|---|---|---|---|
| | Primary eclipses | | |
| 7024 | 58739.9291143 | 1.23 | 1.05 |
| 7025 | 58741.1975064 | 1.23 | 1.48 |
| 7026 | 58742.4659164 | 1.24 | 1.89 |
| 7027 | 58743.7342592 | 1.24 | 2.83 |
| 7028 | 58745.0026685 | 1.23 | 1.40 |
| 7029 | 58746.2710807 | 1.23 | 1.53 |
| 7030 | 58747.5394642 | 1.25 | 1.84 |
| 7031 | 58748.8078501 | 1.24 | 2.02 |
| 7032 | 58750.0762641 | 1.23 | 1.31 |
| 7034 | 58752.6130327 | 1.25 | 1.95 |
| 7035 | 58753.8813901 | 1.23 | 1.87 |
| 7036 | 58755.1498091 | 1.23 | 0.85 |
| 7037 | 58756.4182183 | 1.24 | 1.78 |
| 7038 | 58757.6866029 | 1.25 | 1.57 |
| 7039 | 58758.9549792 | 1.24 | 1.86 |
| 7040 | 58760.2233503 | 1.24 | 0.72 |
| 7041 | 58761.4917561 | 1.23 | 1.45 |
| 7042 | 58762.7601639 | 1.25 | 1.71 |
| Mean | | 1.24 ± 0.01 | 1.62 ± 0.48 |
| | Secondary eclipses | | |
| 7023 | 58739.2936082 | 1.32 | 1.93 |
| 7024 | 58740.5619809 | 1.35 | NaN |
| 7025 | 58741.8303809 | 1.32 | 1.74 |
| 7026 | 58743.0987663 | 1.31 | 1.37 |
| 7027 | 58744.3671520 | 1.32 | 1.93 |
| 7028 | 58745.6355150 | 1.33 | 1.98 |
| 7029 | 58746.9039464 | 1.31 | 1.18 |
| 7030 | 58748.1723268 | 1.31 | 1.11 |
| 7031 | 58749.4406646 | 1.36 | NaN |
| 7033 | 58751.9775236 | 1.31 | 1.58 |
| 7034 | 58753.2458785 | 1.31 | 1.14 |
| 7035 | 58754.5142644 | 1.34 | 2.19 |
| 7036 | 58755.7826500 | 1.32 | 1.75 |
| 7037 | 58757.0510515 | 1.32 | 1.49 |
| 7038 | 58758.3194392 | 1.32 | 1.65 |
| 7039 | 58759.5878394 | 1.32 | 1.86 |
| 7040 | 58760.8562203 | 1.33 | 1.86 |
| 7041 | 58762.1245984 | 1.31 | 1.30 |
| Mean | | 1.32 ± 0.01 | 1.63 ± 0.33 |

**Table A2.** Similar to Table 1, but using the KvW method with seven folds.

| Epoch-Nr. | $T_0$ | $\sigma_{T0}$ | $\sigma_{T0KvW}$ |
|---|---|---|---|
| | (BJD-TBD-2400000) | (10$^{-5}$ d) | (10$^{-5}$ d) |
| | Primary eclipses | | |
| 7024 | 58739.9291150 | 1.28 | 5.49 |
| 7025 | 58741.1975027 | 1.28 | 5.7 |
| 7026 | 58742.4659090 | 1.26 | 2.06 |
| 7027 | 58743.7342707 | 1.26 | 0.88 |
| 7028 | 58745.0026684 | 1.28 | 5.53 |
| 7029 | 58746.2710792 | 1.28 | 5.63 |
| 7030 | 58747.5394538 | 1.26 | 2.38 |
| 7031 | 58748.8078483 | 1.26 | 2.69 |
| 7032 | 58750.0762621 | 1.28 | 5.43 |
| 7034 | 58752.6130182 | 1.26 | 2.23 |
| 7035 | 58753.8813882 | 1.26 | 2.97 |
| 7036 | 58755.1498096 | 1.28 | 5.48 |
| 7037 | 58756.4182146 | 1.29 | 5.67 |
| 7038 | 58757.6865875 | 1.27 | 2.07 |
| 7039 | 58758.9549780 | 1.26 | 2.61 |
| 7040 | 58760.2233496 | 1.29 | 5.51 |
| 7041 | 58761.4917544 | 1.28 | 5.77 |
| 7042 | 58762.7601513 | 1.26 | 1.95 |
| Mean | | 1.27 ± 0.01 | 3.89 ± 1.79 |
| | Secondary eclipses | | |
| 7023 | 58739.2936076 | 1.35 | 2.67 |
| 7024 | 58740.5619515 | 1.35 | 1.62 |
| 7025 | 58741.8303818 | 1.37 | 5.51 |
| 7026 | 58743.0987636 | 1.36 | 5.45 |
| 7027 | 58744.3671522 | 1.35 | 2.27 |
| 7028 | 58745.6355255 | 1.35 | 2.02 |
| 7029 | 58746.9039487 | 1.37 | 5.49 |
| 7030 | 58748.1723265 | 1.36 | 5.42 |
| 7031 | 58749.4407207 | 1.35 | 2.45 |
| 7033 | 58751.9775250 | 1.36 | 5.41 |
| 7034 | 58753.2458815 | 1.37 | 5.32 |
| 7035 | 58754.5142620 | 1.36 | 1.81 |
| 7036 | 58755.7826560 | 1.35 | 2.33 |
| 7037 | 58757.0510536 | 1.37 | 5.65 |
| 7038 | 58758.3194397 | 1.37 | 5.06 |
| 7039 | 58759.5878375 | 1.35 | 2.01 |
| 7040 | 58760.8562260 | 1.35 | 2.09 |
| 7041 | 58762.1245984 | 1.36 | 5.51 |
| Mean | | 1.36 ± 0.01 | 3.78 ± 1.71 |